\documentclass[%
preprint,
 amsmath,amssymb,
 aps,
]{revtex4-2}

\usepackage{graphicx}
\usepackage{dcolumn}
\usepackage{bm}
\usepackage[version=3]{mhchem}
\usepackage{soul,xcolor}
\usepackage{url}

\begin{document}

\preprint{}

\title{Reactive Coarse Grained Force Field for Metal-Organic Frameworks applied to Modeling ZIF-8 Self-Assembly} 

\author{Sangita Mondal$^a$}
\author{Cecilia M. S. Alvares$^b$}
\author{Rocio Semino$^a$}%
 \email{rocio.semino@sorbonne-universite.fr}
\affiliation{$a$) Sorbonne Université, CNRS, Physico-chimie des Electrolytes et Nanosystèmes Interfaciaux, PHENIX, F-75005 Paris, France. \\
$b$) Department of Chemistry, University of Warwick, Coventry CV4 7AL, United Kingdom}%

\date{\today}

\begin{minipage}\textwidth

\begin{abstract}
Decoding the self-assembly mechanism of metal-organic frameworks is a crucial step in reducing trial-and-error tests in their synthesis protocols. Atomistic simulations have proven essential in revealing molecular-level features of MOF nucleation, but they still exhibit limitations in the simulation setups due to size constraints (inability of reaching realistic concentrations or exploring non-stoichiometric metal:ligand ratios). In this contribution, we develop a methodology to derive reactive coarse grained force fields based on multiscale coarse graining methods. We apply our novel methodology to the case of the archetypal zeolitic-imidazolate framework ZIF-8. Our coarse grained force field, which we call nb-CG-ZIF-FF, does not contain any explicit connectivity information, but learns the tetrahedral Zn-connectivity from many body correlations within an atomistic benchmark. nb-CG-ZIF-FF quantitatively reproduces the features of bulk, crystalline ZIF-8 as well as the structural evolution of pre-nucleation species in terms of Zn n-fold coordination populations from the atomistic benchmark. While the range of rings that are formed along the synthesis process are well captured by nb-CG-ZIF-FF, the model cannot exactly reproduce ring populations. Our reactive CG force field fitting approach can be applied to any MOF, opening new research avenues in modeling MOF formation, decomposition, defect dynamics and phase transition processes.  

\end{abstract}

\maketitle
\end{minipage}

\newpage

\section{Introduction}

Metal-organic frameworks (MOFs) are porous materials formed by the combination of metal ions or oxoclusters with organic ligands.\cite{Zhou2012} Since almost any metal of the periodic table and a large collection of polydentate organic ligands can yield MOFs, their pores can adopt a wide range of sizes, shapes and chemical functions. Guest molecules can fill these pores and be stored, separated from other guests or undergo chemical reactions, conferring enormous potential for environmental, medical and industrial applications.\cite{Gatou2023,Yu2024,Chakraborty2023} More than seventy thousand MOFs\cite{Moghadam2017} were synthesized in laboratories all over the world since their discovery by Robson, Kitagawa and Yaghi.\cite{Hoskins1989,Kondo1997,Yaghi1995} But despite this impressive number, our knowledge on the mechanisms that underlie MOF self-assembly is still limited. 

ZIF-8, made from Zn$^{2+}$ cations tetrahedrally coordinated to 2-methylimidazolate (mIm$^-$) ligands,\cite{Park2006} is probably the most widely studied MOF in terms of its synthesis mechanism.\cite{VanVleet2018,Moh2013,Cravillon2012,Lee2015,Pan2011,Feng2016,Bustamante2014,Tsai2016,Beh2018,Malekmohammadi2019,Venna2010,Kida2013,Balog2022,Jin2023,Talosig2024,Dok2025} Based on \textit{in situ} X-ray diffraction and transmission electron microscopy (TEM), Venna and coworkers proposed that an intermediate, amorphous species is formed along the ZIF-8 synthesis process.\cite{Venna2010} More than ten years later, Jin and collaborators could isolate and characterize this amorphous ZIF, which has a slightly reduced Zn-N coordination than the crystalline ZIF-8 ($\sim$3.68 versus 4.0).\cite{Jin2023} Electrospray ionisation mass spectrometry and cryo-TEM allowed Talosig \textit{et al}\cite{Talosig2024} to monitor metal cluster size differences within the amorphous intermediate that were hypothesized to lead to polymorph selection (the porous ZIF-8 MOF or a dense, diamondoid phase). Dok and collaborators have very recently combined harmonic light scattering with NMR to further elucidate the multi-step mechanism of the self-assembly of ZIF-8.\cite{Dok2025}

Computer simulation has been instrumental in revealing molecular-level features of the ZIF-8 synthesis mechanism, along with its solvent- and temperature- dependent dynamics.\cite{Balestra2022,Balestra2023,AndarziGargari2025} Andarzi and Semino modeled the solvothermal nucleation of ZIF-8 in methanol and DMSO\cite{AndarziGargari2025} through an atomistic, partially reactive force field that selectively incorporates Zn--ligand reactivity (nb-ZIF-FF)\cite{Balestra2022}. Their simulations revealed the formation of predominantly linear chains composed by alternating  Zn$^{2+}$  and mIm$^-$ units at early synthesis times. Later on, ramifications and chain merging events lead to the appearance of rings of varied size. The final state of the simulations is not the ZIF-8 crystal, but a highly-connected amorphous aggregate, that could be associated to the amorphous intermediate species discussed above.\cite{Venna2010,Jin2023,Talosig2024,Dok2025} However, the reactant concentrations explored in this study are on the higher end of those experimentally considered. Indeed, these simulations contain $\sim$12000 particles, and studying lower concentrations would require larger systems (and longer timescales), rendering it unfeasible at atomistic resolution. Moreover, assessing the influence of non-stoichiometric metal:ligand ratios in the synthesis process, which are key in determining nanoparticle size and morphology,\cite{Beh2018} would also require systems too large to be treated at the atomistic scale.

Particle-based coarse grained (CG) models, in which groups of atoms are lumped into particles called beads, provide an elegant solution to this problem, as successfully demonstrated by the zeolite research community.\cite{Dhabal2025,ChuJon2025,Bertolazzo2022,Dhabal2021,Bores2022,Bores2018} Within this approach, the number of particles is reduced and their dynamics is accelerated due to the removal of friction forces. Dürholt and coworkers fitted the first CG force field for a MOF by reproducing the Hessian matrix computed from an atomistic benchmark model via a genetic algorithm optimization.\cite{Drholt2016} The authors further used their force field to model mesopore stability\cite{Drholt2016_2} and long-range structural perturbations in MOF/polymer interfaces,\cite{Semino2017} two problems that would have been impossible to tackle at the atomistic level. Alvares and Semino bridged the gap between MOF force field development and CG force field development methods widely used in the polymers and biomolecules communities, such as Multiscale Coarse Graining (MS-CG),\cite{Izvekov2005} Iterative Boltzmann Inversion\cite{Reith2003} and MARTINI\cite{Souza2021}.\cite{Alvares2024} Other CG approaches were proposed to model MOFs, including patchy-particle models,\cite{Scott2025} lattice models,\cite{Gorbunov2023} the micromechanical approach\cite{Rogge2021} and methods specifically tailored for modeling adsorption.\cite{Mohamed2024} While all the CG models described above allow exploring mesoscopic, defective systems, most of them do not allow for modeling synthesis processes, as they do not explicitly incorporate metal-ligand reactivity. Patchy-particles and lattice models have been applied to studying two dimensional self-assembly processes,\cite{Scott2023,Gorbunov2023} but they lack the chemical detail required to model realistic self-assembly processes that include the formation of stochastically generated 3D disordered species.\cite{Scott2025} Doing so requires a chemically specific CG model, tailored to reproduce atomistic interactions, as can be achieved by the MS-CG method.\cite{Koziol2016} In this sense, the MS-CG method originally developed for biomolecules and used to study complex supramolecular processes,\cite{Jin2022,Sahrmann2024} has also been adapted to model SN2 reactions.\cite{DannenhofferLafage2020} In addition, this bottom-up CG force field development approach has led to force fields that reproduce complex chemistry features in MOFs, such as pore gate opening effects,\cite{Alvares2024} which even atomistic models fail to reproduce.\cite{Coudert2017}  

In this contribution, we employ the MS-CG method (also known as force matching) to develop the first reactive CG force field for a MOF: nb-CG-ZIF-FF (non-bonded coarse grained zeolitic-imidazolate framework force field). nb-CG-ZIF-FF is parametrized to reproduce the net forces experienced by the beads computed from atomistic simulations carried out with nb-ZIF-FF\cite{Balestra2022}. It is a fully reactive force field composed by three kinds of beads (Zn, ligand and solvent) that interact through pairwise non-bonded interactions. No point charge nor any kind of bonded contribution are explicitly included in the force field. The tetrahedral character of Zn beads is well reproduced without explicitly including any angle-dependent term in the force field to drive it, contrary to what was done in prior works for zeolites,\cite{Bertolazzo2022,Bores2022} suggesting that topology can be learnt by the MS-CG algorithm. nb-CG-ZIF-FF successfully reproduces both crystalline ZIF-8 structure as well as Zn n-fold coordination profiles along the self-assembly process with remarkable accuracy. Furthermore, relevant structural features found in the atomistic model,\cite{AndarziGargari2025} such as linear chains, rings and the formation of an intermediate amorphous aggregate are all captured by nb-CG-ZIF-FF. All rings that are found in atomistic simulations are also found in the CG case, despite that the relative proportion of ring populations is not reproduced. Our methodology to derive reactive CG force fields for MOFs can be applied to any MOF and even to other porous solids, with a broad impact on the modeling of self-assembly and decomposition processes at experimentally relevant conditions.

This article is organized as follows. Section II summarizes the model, simulation conditions and methodological details. Validation of nb-CG-ZIF-FF with respect to crystalline and self-assembling systems is shown in section III, along with a detailed discussion of our findings. Section IV presents our main conclusions and perspectives.

\section{Methods}

\noindent\textbf{A. System Preparation} \\
We develop a reactive CG force field using the MS-CG algorithm \cite{Izvekov2005} to study the nucleation part of the ZIF-8 self-assembly. To generate reference all-atom (AA) trajectories that serve as the benchmark for our model, we carry out AA simulations using the non-bonded ZIF force field (nb-ZIF-FF). This force field was developed in our group by Balestra and Semino,\cite{Balestra2022} employing cationic dummy atom models to capture the anisotropic electronic density surrounding the metal cation.\cite{Biswal2016} This technique ensures an accurate representation of ZIF-8 topology. nb-ZIF-FF has proven effective in capturing essential properties such as radial distribution function, cell parameters, elastic constant and phase stability in ZIF-8.\cite{Balestra2022} Additionally, it accurately represents the dynamic reactivity between metal ion centers and organic linkers, including bond formation and breaking. Previous studies have demonstrated the success of nb-ZIF-FF,\cite{Balestra2022,Mendez2024,Mendez2025,AndarziGargari2025} making it an ideal starting point for the development of a reactive CG force field.
Here, we consider two distinct molecular systems to parametrize the CG force field: \\
(i) Solvated Zn$^{2+}$ cations and mIm$^-$ anions in dimethyl sulfoxide (DMSO) solvent, for which AA self-assembly trajectories of ZIF-8 formation during the nucleation stage are obtained from previous work in our group by Andarzi and Semino.\cite{AndarziGargari2025}\\
(ii) Solvent loaded crystalline ZIF-8 framework. To saturate ZIF-8 in DMSO, an empty 2×2×2 supercell of the MOF is constructed to serve as the porous host structure to be filled with DMSO molecules. We employ a hybrid Monte Carlo (MC)-molecular dynamics (MD) simulation approach in LAMMPS, \cite{Thompson2022} linked to a high-pressure DMSO reservoir (P=10 atm). At the pressure and temperature selected for the reservoir, DMSO is a stable liquid. The simulation protocol consists of AA MD carried out using Nosé-Hover barostat and thermostat \cite{hoover85} with target pressure and temperature of 1 atm and 298 K, respectively, and a 0.01 fs timestep.  Grand Canonical Monte Carlo (GCMC) moves, featuring molecular insertion and deletion, are attempted every 500 timesteps. The GCMC moves promote reaching the equilibrium adsorption of DMSO molecules within the ZIF-8 pores while the equations of motion used allowed the framework and the adsorbed solvent molecules to structurally relax along the DMSO pore filling process. The simulation is performed until statistical convergence of the DMSO loading is achieved, ensuring that the solvent density inside the framework reflects realistic thermodynamic equilibrium with the external reservoir (see Fig.8). The final system contains 115 DMSO molecules (\textit{i.e.} 14.38 DMSO molecules per ZIF-8 unit cell). To verify that this loading represents true saturation, we perform additional GCMC simulations at reservoir pressures ranging from 10 to 80 atm (see Table II). DMSO loadings remain essentially constant across this pressure range, confirming that the ZIF-8 pores are fully saturated and that our results are independent of the specific reservoir pressure chosen. Subsequently,the system underwent a 30 ns production run using the equations of motion and target P and T previously mentioned. We saved trajectory frames at 25 ps intervals for further processing. Additional details are provided in Supporting Information (SI).\\

\begin{figure}[hbtp!]
\centering 
\includegraphics[width=0.7\linewidth]{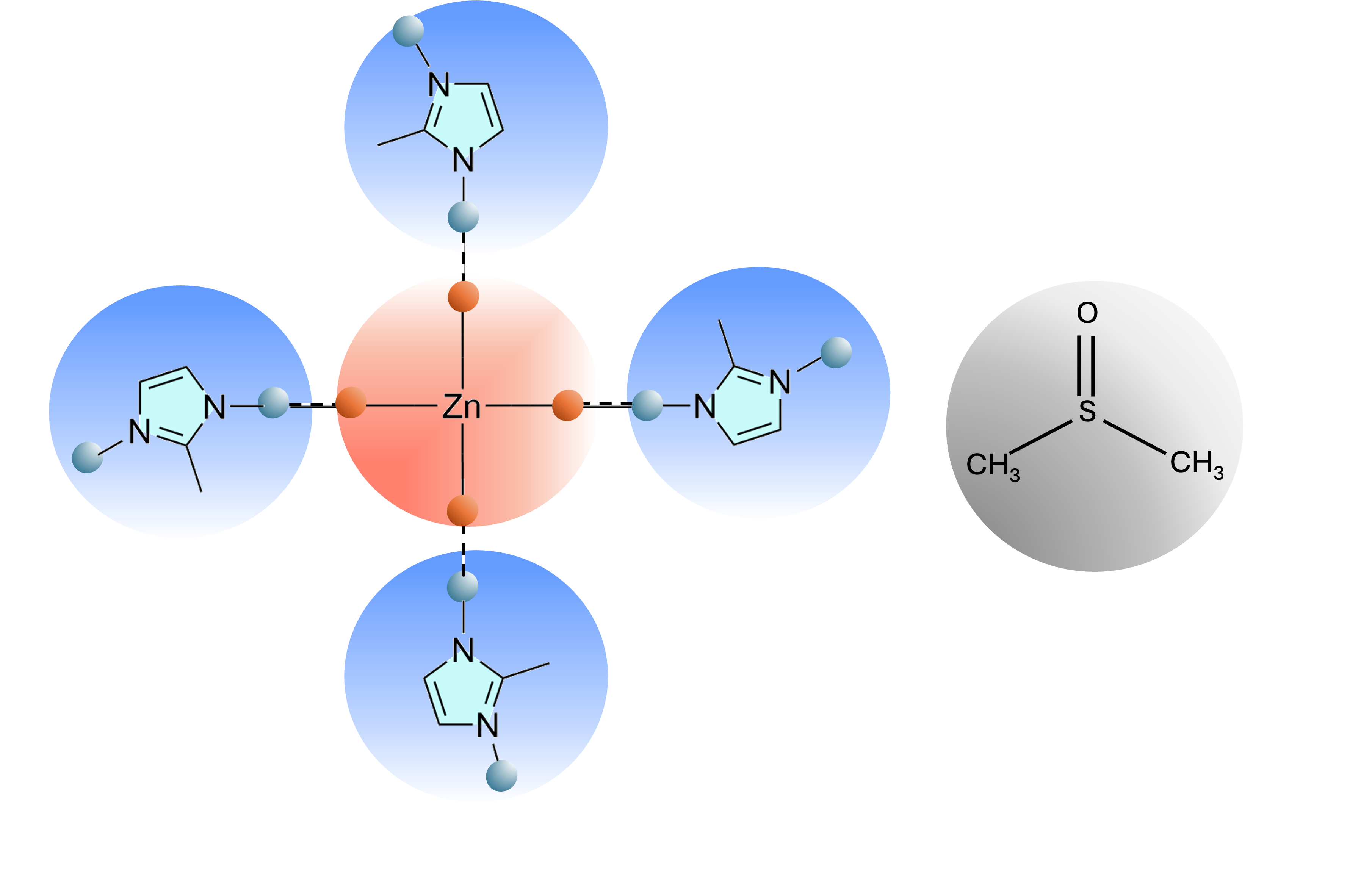}
\caption{Mapping of the coarse grained reactive force field developed in this work: each Zn$^{2+}$ ion together with its four associated dummy atoms[small dark red spheres] is represented as a single bead [light red sphere]; each 2-methylimidazolate (mIm$^-$) ligand along with its two dummy atoms [small light blue spheres] is modeled as an individual bead [dark blue spheres]; and each DMSO molecule is treated as a single bead [gray sphere].}
\label{fig: mapping}
\end{figure}

\noindent\textbf{B. Development of the CG Force Field} \\
The coarse grained representation employs a mapping scheme where each bead corresponds to the center of mass of a designated group of atoms. Multiple AA to CG mapping strategies are possible, ranging from highly coarse (low resolution) to moderately coarse (higher resolution) representations. In a previous computational study by Alvares and Semino,\cite{Alvares2024} it was demonstrated that a two-bead mapping scheme satisfactorily reproduces both structural and thermodynamic properties of ZIF-8, including elastic tensors and thermal expansion coefficients. We follow a similar approach herein, where each Zn$^{2+}$ ion along with its four dummy atoms constitutes a single bead, see Fig.\ref{fig: mapping}.

In the context of the AA benchmark simulations used to parametrize the CG force field, the potential energy function in the AA force field can  be expressed as follows:
\begin{equation}
U_{\text{AA}} = U_{\text{bonded}} + U_{\text{non-bonded}}
\end{equation}
where \( U_{\text{bonded}} \) includes contributions from bond stretching, angle bending and torsional interactions, and \( U_{\text{non-bonded}} \) contains van der Waals (vdW) and electrostatic (charge-mediated) interactions between atoms. Conversely, the CG force field is expressed as:
\begin{equation}
U_{\text{CG}} = U_{\text{non-bonded}}
\end{equation}

\begin{figure*}[h]
\centering 
\includegraphics[width=0.99\linewidth]{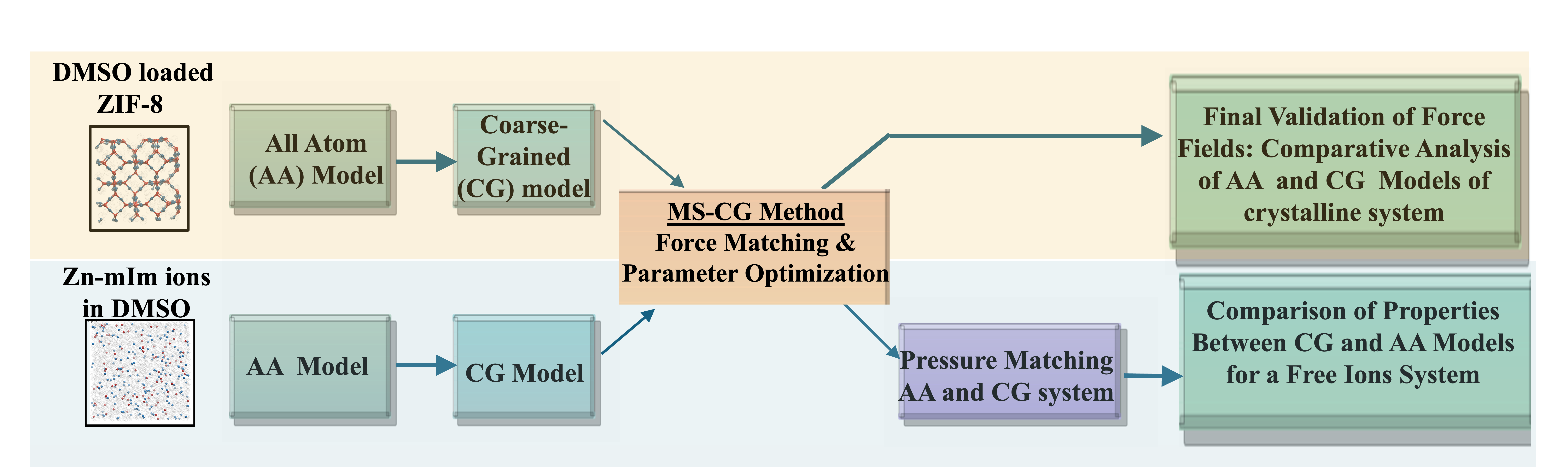}
\caption{ Schematic representation of the CG methodology employed in this work. AA models of crystalline ZIF-8 and the Zn$^{2+}$ and mIm$^{-}$ ions solution in DMSO are converted to CG models and a reactive CG force field is obtained through MS-CG and parameter optimization. Then, a pressure matching stage is performed. The CG force field is validated by comparing observables computed from CG and AA simulations.}
\label{fig:scheme}
\end{figure*}

\text The AA force field includes both bonded and non-bonded terms, while in the CG model, we choose to define the force field as a sum of non-analytical pairwise non-bonded potentials between Zn-Zn, Zn-ligand and ligand-ligand beads, without explicitly enforcing any connectivity. Each of these potentials are given by cubic splines polynomials that hold within an interval of pair distances of 0.005. The cutoff for non-bonded interactions at the CG level is defined as 1.5 nm. 
The complete workflow for developing and validating CG force fields for crystalline ZIF-8 and free ions simulations in DMSO is depicted in Fig. \ref{fig:scheme}.

As shown in the  Fig. \ref{fig:scheme}, following the generation of atomistic reference trajectories we employ the MS-CG technique\cite{Izvekov2005} to systematically derive non-bonded potentials describing interactions between metal, ligand and solvent beads. The MS-CG algorithm is carried out using the BOCS code.\cite{dunn2017bocs} Originally developed by Ercolessi and Adams \cite{ercolessi1994interatomic} for fitting interatomic potentials, the method was adapted to obtaining CG force fields by Izvekov and Voth. \cite{Izvekov2005} The BOCS code subsequently implemented it for CG force field fitting. This method optimizes CG force fields of an arbitrarily pre-established form by minimizing the discrepancy between forces predicted by the CG force field and reference forces obtained from AA simulation trajectories. Here we (i) obtain CG configurations by coarsening the AA configurations following the mapping described in section IIA, and (ii) calculate the reference force experienced by each bead in each CG configuration as the (vectorial) sum of the forces experienced by the atoms that the given bead replaces. This approach enables the generation of physically meaningful and transferable interaction potentials, making it particularly well-suited for complex systems such as MOFs. \cite{Alvares2025,Alvares2024} An in-depth discussion of the force matching procedure within the MS-CG method can be found on the SI of Refs. \citenum{Alvares2025} and \citenum{ruhle2009}. In this study, we introduce a dual-trajectory approach that combines simulations for the two molecular systems mentioned in section IIA to parametrize the CG force field: a trajectory of a crystalline ZIF-8 filled with DMSO and three other independent trajectories initially composed by 96 Zn$^{2+}$ and 192 mIm$^-$ ions solvated in 2344 DMSO molecules. Specifically, we employ the crystalline ZIF-8 trajectory to provide equilibrium structural data for the crystalline phase, alongside with three ionic solution trajectories coming from independent simulations that capture the stochastic nature of dynamic interactions between metal ions and ligands, including association and dissociation events. This combination is intended to ensure that the force field accurately represents both the equilibrium properties of the crystalline structure and the diverse pathways of ionic self-assembly.
BOCS allows training over a series of reference trajectories with user-defined weights for each.\cite{dunn2017bocs} We choose a composition ratio of 0.7 crystalline ZIF-8 trajectory to 0.3 for the self-assembly trajectories (0.1 for each independent trajectory). Note that the proportions of the trajectories influence the resulting non-bonded interactions, as discussed below.
By applying these ratios to the training system, we obtain a comprehensive configurational sampling that encompasses both crystalline ZIF-8 environments and all the diverse species that are generated during the self-assembly process. We stress that despite the fact that the individual pairwise potentials defining the CG force field are isotropic by construction, the forces predicted by the MS-CG force field for a given bead in a given configuration are the result of a combined effect of all potentials, thus allowing for anisotropy to still occur. \\

Following the MS-CG calculations, the forces as a function of distances between all pairs of beads are smoothened to remove numerical noise. Due to the form assumed for the CG force field (i.e. given by non-analytical pair potentials with small bin sizes), the MS-CG algorithm only outputs data for pair distances that are sampled within the trajectory used as input. Thus, the force profiles do not contain data for pair distances that do not occur during the dynamics. Radial distribution functions built using the coarsened AA trajectories served as essential guides for determining the threshold for which sampling exists. The pair potential is fitted using a cubic spline. The spline was smoothly extended into the non-sampled short-range region to ensure continuity of the potential and its derivatives, following standard practice in spline-based interatomic and pair-potential formulations. \cite{xie2023ultra,lenosky2000highly,stoller2016impact} \par
We note that developing a force field that simultaneously stabilizes the crystalline ZIF-8 structure while capturing realistic self-assembly dynamics proved challenging and required extensive manual refinement of the weighting considered for the trajectories. Potentials optimized for crystalline stability often produced overly rigid interactions that hindered assembly kinetics, whereas potentials favoring dynamic solution-phase behavior sometimes compromised lattice integrity. Through iterative adjustments and systematic testing against both structural metrics (lattice parameters and radial distribution functions) and assembly dynamics ( cluster formation), we obtain a balanced force field capable of accurately describing both the stable crystalline framework and solution-phase self-assembly processes. The fitted and raw (noisy) force–distance curves together with further details on the fitting procedure are provided in the SI(see Fig.9).
After establishing the potentials, CG simulations starting from Zn$^{2+}$ and mIm$^-$ solutions in DMSO are conducted under the NVT ensemble with temperature 298 K for validation. However, during the self-assembly process, the system density varies significantly, making a constant-volume ensemble not optimal for capturing realistic dynamics.\cite{quigley2009metadynamics} To circumvent this issue and maintain the structural stability of the MOF framework during equilibration, we implement the \emph{fix deform} algorithm available in LAMMPS.\cite{lammps} This approach enables controlled modification of the simulation box dimensions, allowing the system to gradually relax toward the expected equilibrium volume by the end of the self-assembly process. Expected volumes are calculated from the final states observed in the AA model.

For more robust and realistic validation of the force field, we subsequently perform simulations using Nosé-Hoover thermostat and barostat,\cite{hoover85} with constant temperature and pressure T= 298 K and P=1 atm and damping constants of 100 fs and 1000 fs respectively.  However, these runs exhibited severe instabilities in volume, resulting in nonphysical expansions of the simulation box.
This phenomenon arises from fundamental limitations inherent to CG models in reproducing pressure.
In molecular simulations, the instantaneous pressure is calculated using the virial equation:\cite{Martyna1999,tuckerman2023statistical}

\begin{equation}
P = \frac{1}{dV} \left[ \sum_{i=1}^{N} \frac {\mathbf{p_i^2}}{m_i} + \sum_{i=1}^{N} \mathbf{r_i} \cdot \mathbf{f_i} - (dV)\left( \frac{\partial U(\mathbf{r},V)}{\partial V} \right) \right]
\end{equation}

\noindent
where \(\mathbf{p}_i\) refers to the momentum vector of the $i$-th particle in the system, \({m}_i\) represents the mass of the system, \(V\) is volume, \(d\) is dimensionality (\(d=3\)), \(\mathbf{r}_i\) is the position vector of particle \(i\), \(U(\mathbf{r},V)\) is the potential energy and \(\mathbf{f}_i\) is the total force acting on particle \(i\). The first term represents the kinetic (ideal gas) contribution, while the second term represents the virial contribution from interparticle interactions and the third term accounts for the volume dependent part of potential energy. In the pressure equation convention adopted by LAMMPS,\cite{lammps} the volume-dependent contribution is not included.
In CG models, reduction of degrees of freedom does not allow the virial and kinetic terms of the pressure calculation to reproduce the AA counterpart at a given thermodynamic state. Without explicit pressure correction terms, such as those proposed by Das and Andersen \cite{das2010multiscale} and Izvekov and Voth, \cite{Izvekov2005} CG potentials usually fail to reproduce the correct virial and consequently yield inaccurate pressure predictions. To overcome this limitation, we follow the iterative pressure matching technique by adding a volume-dependent potential, U$_{v}$(V), into the Hamiltonian.\cite{hill2012introduction} This term represents the system contribution to the thermodynamic pressure as it can be seen from equation (3) for thermodynamic pressure derived within statistical mechanics. The underlying idea of pressure matching is to fit a U$_{v}$(V) term for the force field whose partial derivative \(\Bigl(\frac{\partial{U_{v}(V)}}{\partial V}\Bigl)\) allows reproducing the pressure at the (T,V) condition of choice in the context of equation (3). These terms are fit for the Zn$^{2+}$ and mIm$^-$ solution in DMSO system to enable reproducing the pressure at the (T,V) condition where ambient pressure is expected. We note that this is the same state of the reference AA simulations that are employed in the fitting of the MS-CG potentials.

We make the pressure correction using the mathematical framework from Das and Andersen \cite{das2010multiscale}:

\begin{equation}
U_v(V) = \psi_1 N \frac{V}{\bar{v}} + \sum_{i=2}^{n} \psi_i N(\frac{V - \bar{v}}{\bar{v}})^i
\end{equation}

 In this expression, $N$ denotes the total count of CG beads, $n$ stands for the number of atoms in the reference AA system, while $\bar{v}$ and $V$ represent the mean volume obtained from AA and CG simulations under the specific thermodynamic state being modeled, respectively. The parameter values for this expression are obtained through the BOCS code.\cite{dunn2017bocs}
The procedure begins with an initial AA trajectory and a corresponding CG trajectory. Using these, we perform a pressure matching step to obtain starting values for the parameters $\Psi_1$ and $\Psi_2$. These parameters are then used to run a new CG simulation. If the resulting pressure and volume do not match the target values obtained from AA trajectories, we perform pressure matching again to refine $\Psi_1$ and $\Psi_2$, and use the updated parameters to run another CG simulation. This iterative loop continues, adjusting the pressure-matching parameters after each CG run, until the pressure matches the AA one. Note that for crystalline ZIF-8 no pressure correction is applied, as the average pressure and volume are already consistent with the all-atom reference values. Pressure correction is only applied to study self-assembly.\\

\noindent\textbf{C. CG Simulations Details} \\

Once the reactive CG force fields are obtained, we evaluate their ability to reproduce structural properties, such as radial distribution functions (RDFs) and cell parameters, as well as to model self-assembly phenomena in solvated metal--ligand systems. To validate the nb-CG-ZIF-FF in its capacity for modeling nucleation processes, we compute coordination and ring distribution evolution analyses. Coordination number (CN) is defined as the number of ligand beads within 3.6 \AA \ of each metal bead, corresponding to the first minimum in the metal-ligand RDF. Metal beads are classified by coordination state (CN = 0–4) and populations are monitored throughout the simulation. We also perform ring analyses to identify closed metal-ligand cycles of size 3–8 metals, where rings are defined as cyclic paths of alternating metal centers and bridging ligands with each ligand coordinating exactly two metals in the ring. Ring statistics provide direct insight into the emergence of ZIF-8's characteristic sodalite cage topology.\cite{Park2006} Detailed explanations of the algorithms for both analyses are provided in the Supporting Information.
Moreover the results are compared with the respective reference AA systems.
All molecular dynamics simulations are performed using LAMMPS.\cite{lammps} The initial configurations are energy-minimized using the conjugate gradient algorithm until the maximum force falls below 10$^{-4}$ kcal/mol/\AA. From there, two simulation setups are considered: (i) simulations where the simulation box is deformed with the LAMMPS \emph{fix deform} command at a rate of 2.5×10$^{-7}$ nm/simulation step for all simulation box dimensions. The box deformation is coupled with dynamics where a Nosé-Hoover thermostat \cite{hoover85} is deployed to maintain a constant temperature of 298 K. Production simulations are conducted with a timestep of 1 fs and system trajectories are dumped at regular intervals of  0.2 ps for further analysis. (ii) NPT ensemble simulations for both crystalline ZIF-8 and the self-assembly system at T = 298 K and P = 1 bar. Temperature and pressure are controlled by Nosé-Hoover thermostats and barostats  \cite{Evans1985} with damping times of 100× and 1000× the timestep, respectively. For each setup, production runs of 1 ns are performed. For the self-assembly system, we generate three independent trajectories starting from different initial conditions to capture its stochastic nature. It should be noted that since the definition of pressure in LAMMPS does not include the volume-dependent contribution, the study of pressure-corrected models under NPT conditions is performed using the \emph{fix bocs} command to account for volume-dependent Hamiltonians\cite{lammps_npt,lammps_bocs,dunn2017bocs,Martyna1994constant} and these pressure-correction is specifically employed for the self-assembly simulations.

Interestingly, our MS-CG approach shares conceptual similarities with machine learning force fields (MLFF), particularly in that both imply data-driven parameterization.\cite{Behler2007} Both learn interaction potentials directly from simulation data rather than empirically  fitting analytical, physically-derived functions. MLFFs use neural networks trained on expensive DFT data to achieve quantum accuracy at the atomistic scale for equilibrium properties, while the MS-CG technique uses pairwise potentials fitted from classical MD trajectories. Together, they represent a paradigm shift from empirical force field design toward systematic, data-driven approaches that can bridge quantum mechanics and mesoscale phenomena in complex self-assembling systems.\cite{tayfuroglu2026transforming,krass2025mofsimbench,castel2024machine,dobbelaere2025cluster}.

\section{Results}

\noindent\textbf{A. nb-CG-ZIF-FF force field validation: crystalline ZIF-8}\\ 

 We first validate nb-CG-ZIF-FF for its ability to reproduce structural features of crystalline ZIF-8.
 \begin{figure}[hbtp!]
\includegraphics[width=0.9\linewidth]{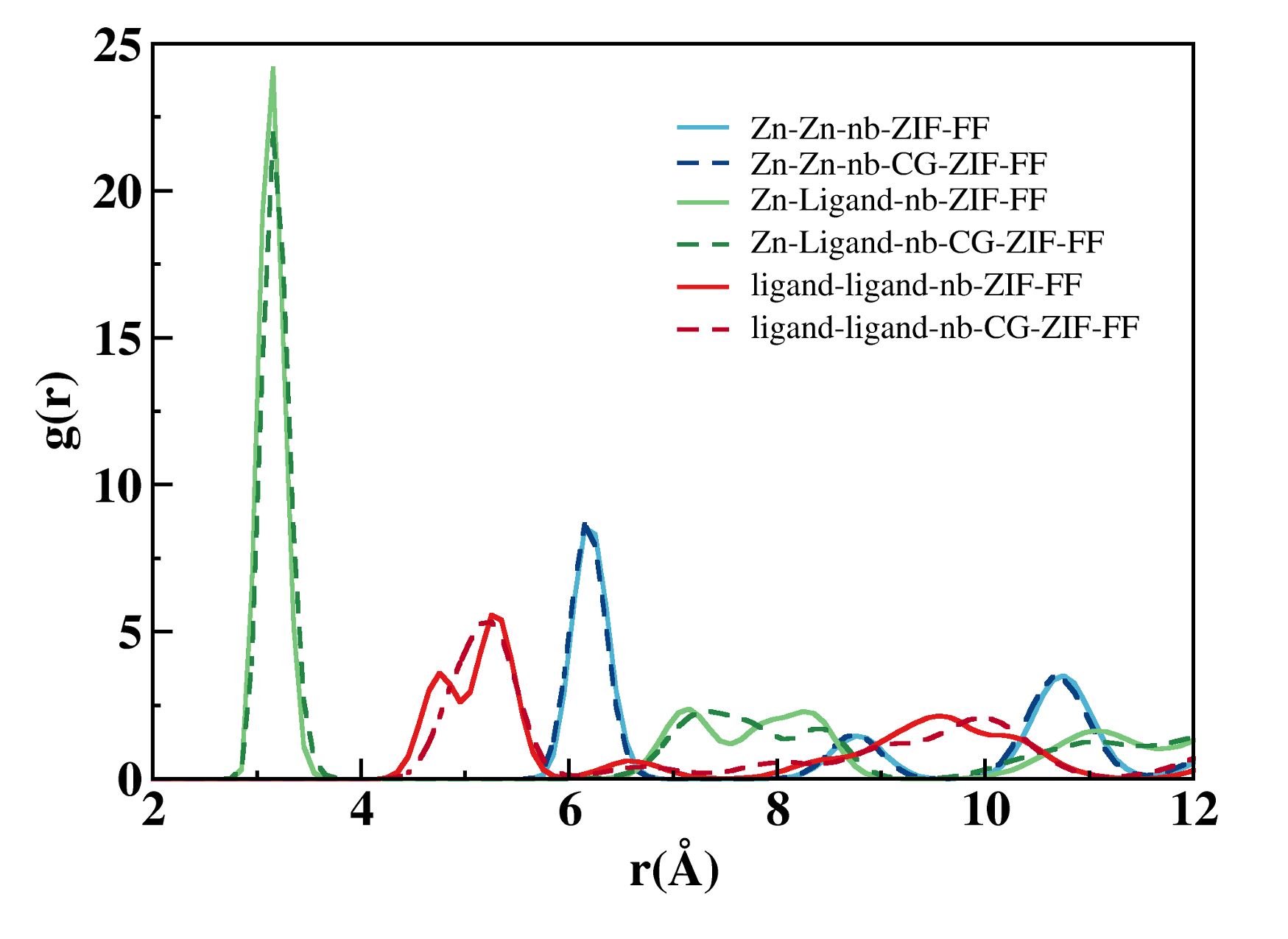}
\caption{Radial distribution functions g(r) for Zn-Zn (blue), Zn-ligand (green), and ligand-ligand (red) interactions in crystalline ZIF-8, obtained from the nb-CG-ZIF-FF model  (dashed lines) and from coarsening the AA trajectory (solid lines).}
\label{fig:rdf}
\end{figure}
 Fig. \ref{fig:rdf} shows RDFs for all bead-bead pairs in the MOF. Solid lines represent the RDFs obtained from the coarsened AA trajectories that serve as the reference target. Dashed lines represent the RDFs generated from simulations using our developed nb-CG-ZIF-FF model (without pressure correction). Three types of bead-pair interactions are examined: Zn-Zn (blue), Zn-ligand (green) and ligand-ligand (red). The most prominent structural feature in the figure is the Zn-ligand interaction, which exhibits a very sharp and intense peak around 3.15 \AA, given by beads that are adjacent to each other, thus representing the coordination bond. The Zn-Zn interaction displays a distinct peak around 6.16 \AA, while the ligand-ligand interaction shows broader peaks in the 4-5 \AA \ region. Notably, not only the first peaks but also the second peaks and subsequent structural features at larger distances are well captured by nb-CG-ZIF-FF, as proven by the excellent agreement between the dashed and solid lines for all interaction types. Additionally, the lattice constant is well reproduced, as shown in Table \ref{tab:table2}, which presents the cell parameters computed via nb-CG-ZIF-FF (ZIF-8 filled with DMSO), the AA reference system (ZIF-8 filled with DMSO) and from nb-ZIF-FF calculations made on crystalline ZIF-8 without solvent (taken from Balestra and Semino \cite{Balestra2022}). 
 
 \begin{table}[h]
\caption{\label{tab:table2}
Cell parameters of empty ZIF-8, AA-DMSO loaded ZIF-8 and nb-CG-ZIF-FF-DMSO loaded ZIF-8. }
\begin{ruledtabular}
\begin{tabular}{cccccccc}
 &$a$ (\AA)&$b$ (\AA)&$c$ (\AA) \\
\hline
CG DMSO-loaded ZIF-8 & 17.41 & 17.41 & 17.41 \\
AA DMSO-loaded ZIF-8 & 17.50 & 17.50 & 17.50 \\
AA empty ZIF-8\footnotemark[1]& 16.99 & 16.99 & 16.99 \\
\end{tabular}
\end{ruledtabular}
\footnotetext[1]{{Taken from, Ref.~\citenum{Balestra2022}}}
\end{table}
It is important to mention that the cell parameter of ZIF-8 is larger in solvent-loaded systems. This expansion is expected, and it occurs due to the adsorption of DMSO molecules within the framework pores, which exert internal pressure on the ZIF-8 structure, causing the lattice to expand to accommodate the guest molecules as experimentally observed.\cite{hobday2018understanding} We can thus conclude that our force field successfully reproduces the structural properties of the atomistic ZIF-8 framework. \\

\noindent\textbf{B. nb-CG-ZIF-FF force field validation: self-assembly}\\
Next, we subject nb-CG-ZIF-FF to a more stringent test: to see whether it can reliably model ZIF-8 nucleation. To this end, we perform MD simulations with constant number of particles, volume and temperature starting from a configuration containing Zn and ligand beads fully solvated by DMSO beads, as described in the methodology section, and we let the system evolve.
\begin{figure*}[h]
\includegraphics[width=0.7\linewidth]{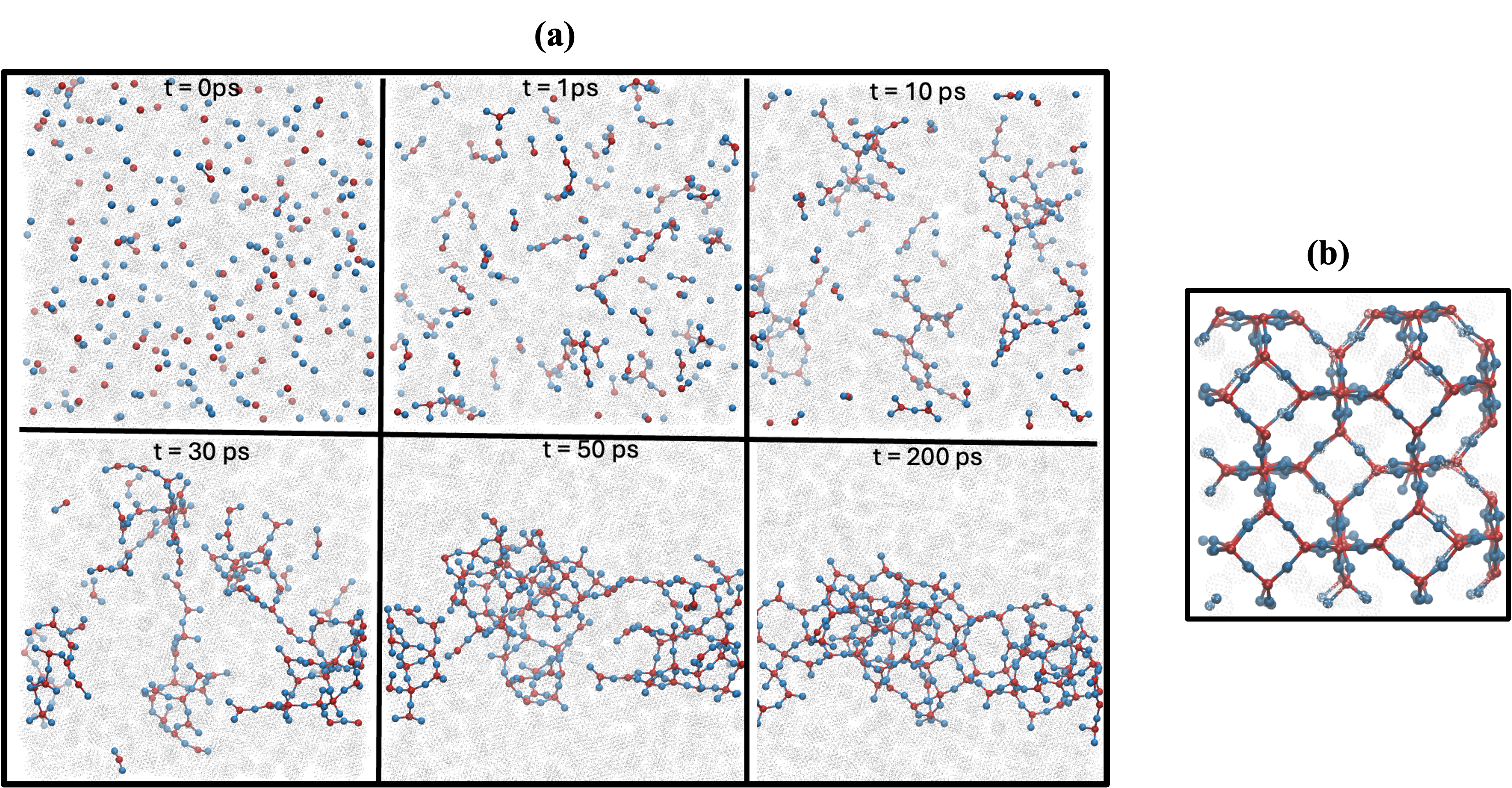}
\caption{ Snapshots of a molecular dynamics simulation of ZIF-8 self-assembly made with nb-CG-ZIF-FF. (a) Sequential snapshots (t = 0–200 ps) showing the progression from dissolved metal and ligand beads through the formation of linear chains, branched oligomers, and interconnected networks, up to an amorphous ZIF-8 structure. (b)  View of a representative crystalline ZIF-8 structure at the CG mapping chosen in this work.}
\label{fig:snap}
\end{figure*}

In Fig.~\ref{fig:snap} simulation snapshots from the \textit{fix deform} system at different times reveal a hierarchical self-assembly mechanism: at $t = 0$ ps, the system contains dispersed metal and ligand precursors. Within the first picosecond, rapid metal--ligand coordination generates small, mostly linear chains, which subsequently extend into larger oligomeric chains. As assembly progresses ($t = 10$--30 ps), these chains extend and branch, forming cyclic motifs. Between $t = 50$--100 ps, continued polymerization and merging of these ring-containing oligomers produce an interconnected three-dimensional network creating cage-like structures that ultimately aggregate into amorphous ZIF-8 networks. Similar behavior is also observed in simulations performed under constant NPT conditions. This computationally observed pathway has been previously depicted in atomistic simulations, \cite{Balestra2022,AndarziGargari2025} and it provides direct molecular-level visualization of the nucleation process. Our CG simulations successfully reproduce all qualitative features that are found in the AA systems, demonstrating for the first time that the nucleation mechanism can be captured at reduced resolution with significantly improved computational efficiency, yielding speeds up of 2 orders of magnitude. In particular, the formation of the final stable amorphous species, characterized by a 4-fold coordinated Zn population of approximately 60\%, required roughly 15 days of wall-clock time in AA simulations, whereas the same structure is achieved within about 2 hours using the CG model, while preserving the same qualitative nucleation pathway. Furthermore, these simulation findings are strongly supported by experimental studies demonstrating the crucial role of amorphous intermediates in ZIF-8 formation.\cite{Jin2023,Talosig2024,Venna2010} Our CG simulations capture the initial stages of this process, revealing how chain formation, extension, and ring closure lead to the experimentally observed pre-nucleation clusters and amorphous assemblies.

\begin{figure}[h]
 \centering 
\includegraphics[width=0.66\linewidth]{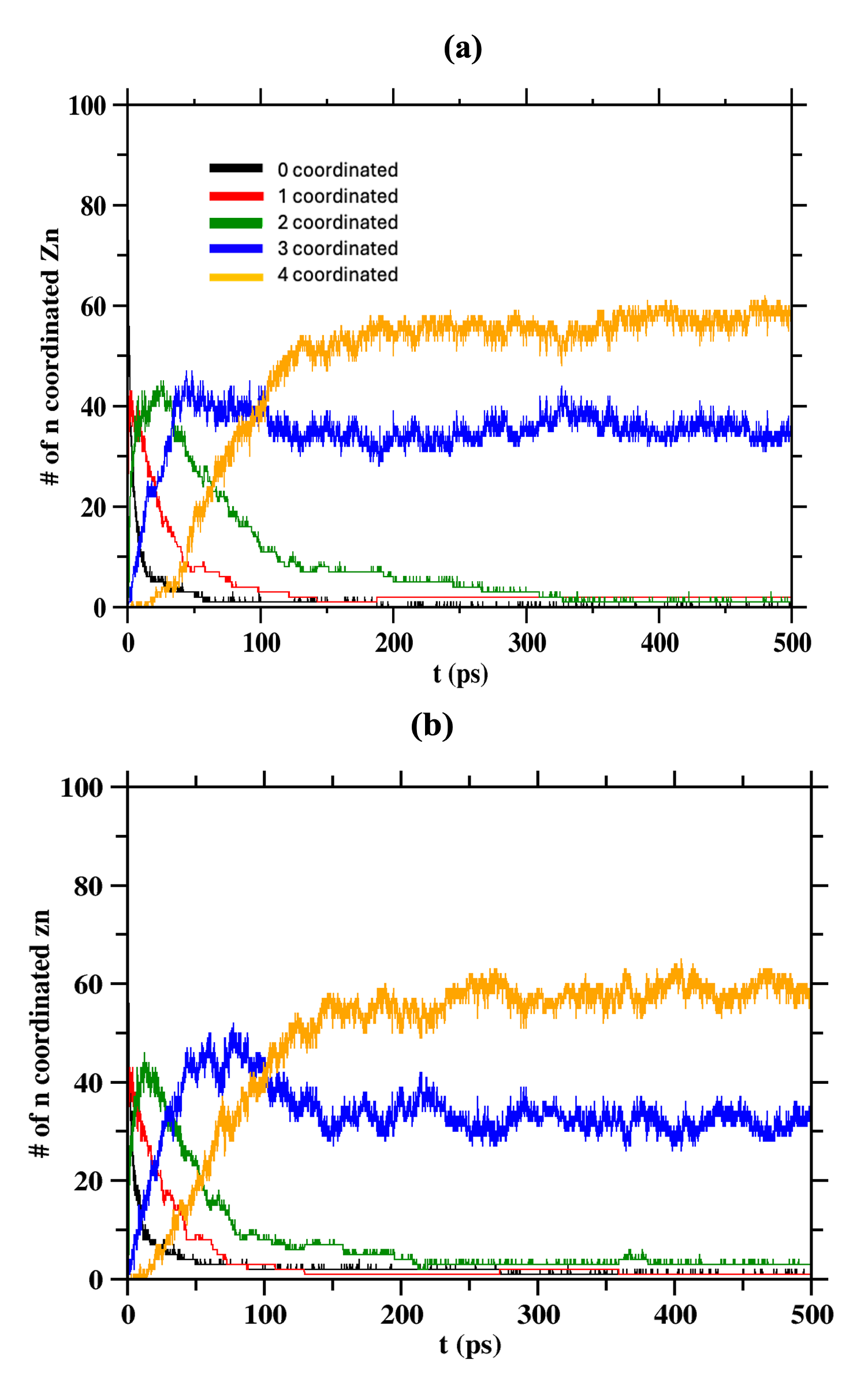}
\caption{Zn-ligand coordination number evolution: (a) NVT simulation  (b) NVT simulation carried out with the \emph{fix deform} command, where the system volume is allowed to gradually relax to match the final volume from the reference AA NPT simulation.}
\label{fig:nvt_fix_deform}
\end{figure}

In order to study the capability of nb-ZIF-FF in modeling the self-assembly in a more quantitative way, we analyze the time evolution of the coordination number between metal and ligand beads. Fig.\ref{fig:nvt_fix_deform} (a) and (b) compare the n-fold coordination numbers obtained from the NVT and NVT \emph{fix deform} trajectories discussed in the methodology section.

We can see in the figure that the population of 0-coordinated Zn beads (black curve) decreases rapidly within the first 50 ps. The fractions of 1- and 2-coordinated (red and green) Zn species exhibit a sharp initial rise at the early stage, reaching their respective peaks before gradually declining. In contrast, the 3-coordinated Zn beads (orange curve) increase more slowly and eventually saturate at 35-40\%. The 4-coordinated Zn species emerge later in the process, showing a steady increase and reaching a plateau after 100 ps. In the long run, the system is dominated by 4-coordinated Zn beads (60\%), followed by a smaller fraction of 3-coordinated species, with very few other coordination variants remaining. These simulations reproduce remarkably well overall profiles of the time evolution of n-fold coordination numbers observed by Andarzi and Semino\cite{AndarziGargari2025} for the AA reference system, albeit within a reduced timescale, as expected in CG simulations.

\begin{figure*}[h]
\centering 
\includegraphics[width=0.7\linewidth]{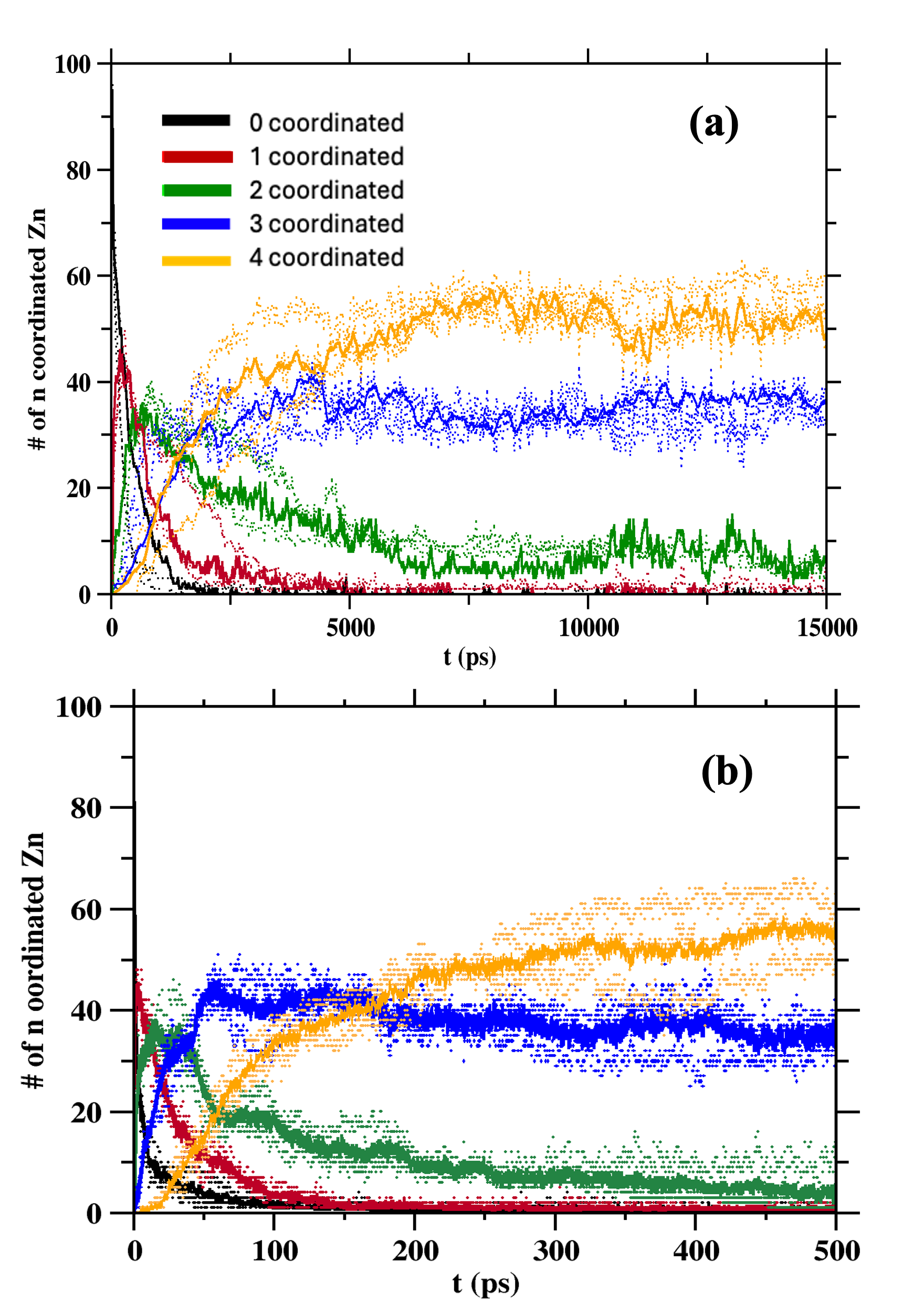}
\caption{ Coordination dynamics of Zn$^{2+}$ ions and Zn beads during ZIF-8 self-assembly from (a) AA and (b) CG simulations respectively. Dotted lines indicate the data obtained from independent parallel simulation trajectories.}
\label{fig:n fold coordination}
\end{figure*}

Subsequently, we performed NPT simulations using the pressure corrected force field instead of using the force field without correction coupled to the explicit box deformation approach. Panel (a) of Fig.~\ref{fig:n fold coordination} shows the results obtained from the AA reference trajectories, while panel (b) presents the corresponding data from simulations using the nb-CG-ZIF-FF model. A comparison between these two datasets allows us to assess the accuracy of the CG force field. Both AA and CG curves display consistent yet temporally shifted behavior across all possible Zn bead coordinations. In the 0- and 1-fold states, the AA model displays slow dynamics, with populations either collapsing rapidly (0-fold) or forming transiently before decaying on nanosecond timescales (1-fold). In contrast, the CG model responds almost instantaneously, producing sharp early peaks that relax within hundreds of picoseconds. In the 2-fold coordinated curve (green), the AA model shows a slow but steady buildup over a few nanoseconds that holds on for a while. The CG model, on the other hand, peaks within the first hundred picoseconds and then drops back rapidly. Moving to the 3-fold curve (blue), the AA curve slowly builds up over a couple of nanoseconds, while the CG counterpart peaks early and then gradually comes down and plateaus. Interestingly, by the end both models end up with a final population of tri-coordinated Zn of 40\%. Both models reach a final population of around 60\% for the 4-fold state. The AA model rises slowly and steadily (reaches in 5 ns), while the CG model plateaus early and then climbs up gradually (200 ps).

The three simulation methods (NVT, NVT with fix deform, and pressure-corrected NPT) produce similar results. As expected, for all three simulation methods (\textit{i.e.} fixed volume, explicit box deformation via \textit{fix deform} or NPT with the pressure correction term), CG simulations exhibit faster kinetics than AA simulations do due to the reduced number of degrees of freedom. However, it is remarkable that the CG model not only accelerates the dynamics but also successfully reproduces the overall shape and characteristic features of the AA curves, capturing the essence of the nucleation process.
We stress that in our CG model, no explicit potential form is considered to ensure the tetrahedral coordination environment of the Zn$^{2+}$ centers. Nevertheless, the simulations clearly show that 4-fold coordinated Zn species are predominant. This observation suggests that the CG model implicitly learned the tetrahedral coordination tendency from the AA reference trajectories during the force-matching procedure and successfully reproduced this structural feature in the CG simulations. This emergent tetrahedrality is particularly noteworthy when compared to explicitly parameterized approaches in zeolite synthesis modeling. Bertolazzo \textit{et al.}\cite{Bertolazzo2022} and Dhabal \textit{et al.}\cite{Dhabal2021} developed coarse grained models for zeolite formation where silica is represented as a single tetrahedral network former particle (denoted as T). In these models, tetrahedral coordination was explicitly enforced through a polynomial three-body interaction term designed to favor the T-T-T (silica-silica-silica) angles characteristic of silica networks. The interactions combine two-body and three-body terms to favor silica angles, enabling the study of polymerization and crystallization processes in zeolite synthesis. In contrast to these explicit tetrahedron inducing strategies, nb-CG-ZIF-FF derives its tetrahedral preference implicitly from the many-body correlations present in the atomistic training data, suggesting that force-matching can effectively encode geometric constraints without requiring their manual specification in the functional form of the potential.

\begin{figure}[hbtp]
 \centering 
\includegraphics[width=0.6\linewidth]{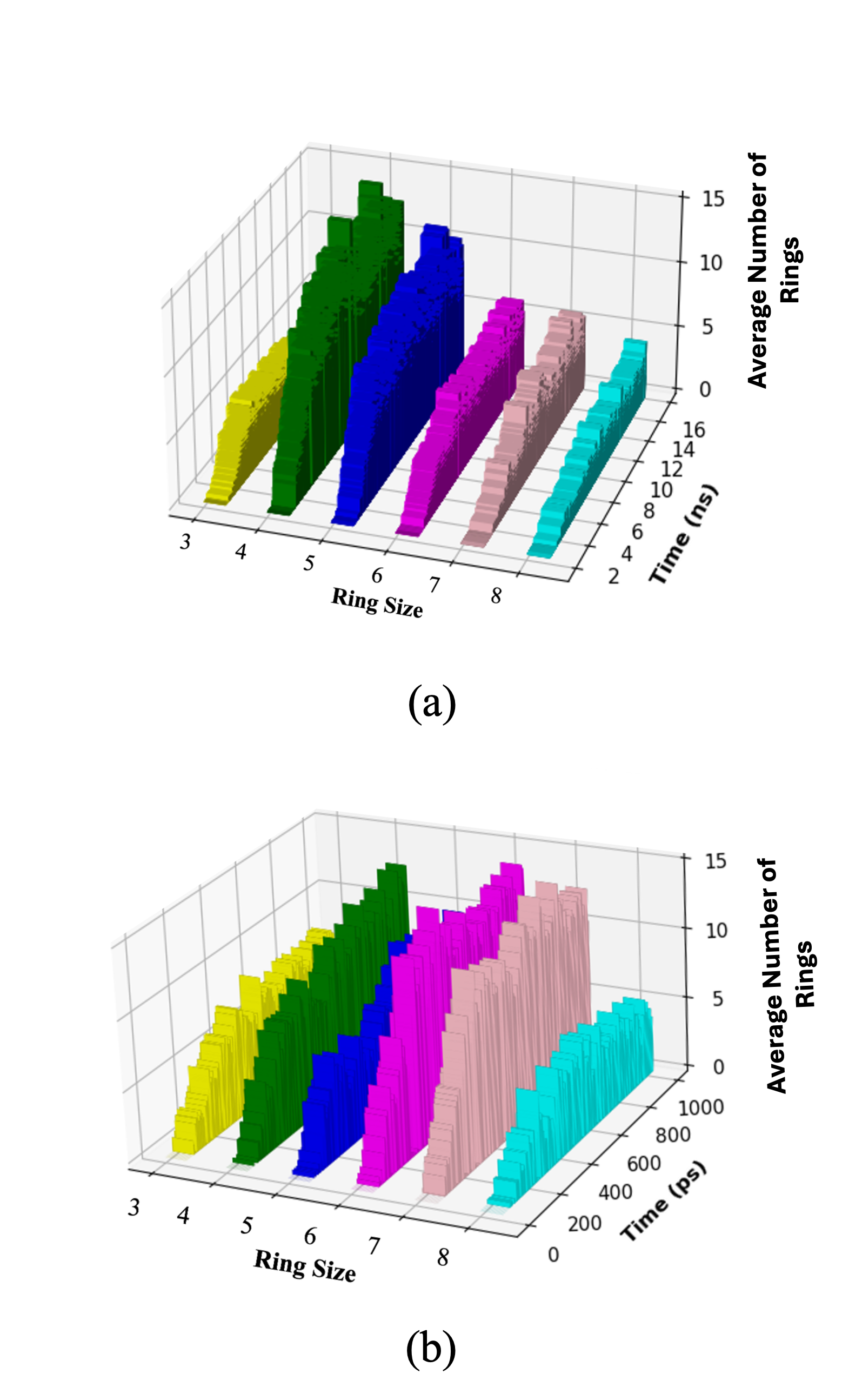}
\caption{Time evolution of ring size distributions during the simulated ZIF-8 self-assembly process. Results from (a) AA reference simulations and (b) nb-CG-ZIF-FF simulations. Data represent averages over three independent simulation trajectories.}
\label{fig:ring}
\end{figure}

Finally, we go deeper into investigating whether our force field captures detailed structural features of the self-assembly process, to discover its limitations. CG force fields always have limitations with respect to the information we can obtain from them, as the reduction of degrees of freedom implies the loss of part of the chemical nature of the system. With this in mind, we analyze the structure and dynamics of ring formation, which is expected to be strongly correlated to the chemistry of the ligands. Indeed, ring size distribution is a highly sensitive property, as it depends on the precise formation and rearrangement of coordination bonds. Although ZIF-8 is known to contain only 4- and 6-membered rings, the formation of 3-, 5-, 7-, and 8-membered rings has been described for amorphous ZIF-4\cite{Castel2024,liu2025amorphous,mendez2024phase} and it has also been predicted for the amorphous intermediate of ZIF-8 formed in the work by Andarzi and Semino that we have used as a reference herein.\cite{AndarziGargari2025} The upper panel of Fig.~\ref{fig:ring}  shows the ring distribution obtained from the AA trajectories, while the lower panel presents the corresponding results from the nb-CG-ZIF-FF model

Analysis of the AA trajectories shows that 4-membered rings are the most abundant, followed by 5-, 6-, and 7-membered rings. The nb-CG-ZIF-FF simulations predominantly feature 4-, 6-, and 7-membered rings, closely reproducing, though not matching, the trends observed in the AA simulations. Despite the inherent loss of degrees of freedom, the nb-CG-ZIF-FF model reproduces the ring distribution reasonably well, although not capturing the global relative ring populations trend. This highlights the robustness of the coarse grained representation in reflecting the essential structural characteristics of the system.

\section{Conclusions}

In this work, we develop a reactive CG force field (nb-CG-ZIF-FF) to model how ZIF-8 assembles from its metal and ligand building blocks in a DMSO solution. Our force field is developed through the MS-CG method,  incorporating a pressure correction via pressure matching for self-assembly studies at constant temperature and pressure. nb-CG-ZIF-FF accurately reproduces both the structural properties of crystalline ZIF-8 and the dynamics of its self-assembly process from an atomistic benchmark, while achieving computational efficiency gains of $\sim$2 orders of magnitude.

nb-CG-ZIF-FF successfully captures the essential structural features of crystalline ZIF-8, including short- and long-range Zn-ligand coordination modes. Validation against all-atom reference simulations demonstrates quantitative agreement in radial distribution functions across all bead-pair interactions. Most excitingly, our simulations successfully reproduce the step-by-step assembly process that was previously seen in all-atom simulations \cite{Balestra2022,AndarziGargari2025} in terms of the time-evolution profiles of n-coordinated Zn populations (n=0-4). This represents the first demonstration that MOF nucleation dynamics can be reliably captured at CG resolution while reasonably preserving chemical accuracy.
A particularly notable feature of our nb-CG-ZIF-FF force field is its ability to implicitly capture the tetrahedral coordination preference of Zn$^{2+}$ ions without explicit angular potentials or geometric constraints. Our force field encodes these geometric preferences implicitly through the combined effect of the CG potentials in reproducing reference forces. This suggests a broader lesson: for systems where local coordination environments emerge from chemical interactions rather than requiring hard constraints, data-driven parametrization may offer greater flexibility  than functional forms designed for specific geometries. As the system evolves, we see it go from scattered building blocks to coordinated chains and eventually to an amorphous network where about 60\% of the Zn atoms have their preferred 4-fold coordination, with another 35\% having 3-fold coordination, quantitatively matching the steady-state reached in the benchmark AA simulations. Our simulations reach an amorphous state, consistent with amorphous intermediates that experimentalists have isolated and characterized prior to ZIF-8 crystallization.
 Analysis of ring formation kinetics reveals that the nb-CG-ZIF-FF model captures the emergence of 4-, 5-, 6-, 7- and 8-membered rings during network growth, consistent with both all-atom simulations and experimental characterizations of amorphous ZIF phases. \cite{AndarziGargari2025,Castel2024,Venna2010,Talosig2024,Dok2025,liu2025amorphous,mendez2024phase} While the CG representation necessarily sacrifices atomistic detail, the model preserves the essential topological features governing framework assembly, albeit without perfectly capturing ring population tendencies. The accelerated dynamics observed in CG simulations as a consequence of reducing the number of degrees of freedom does not compromise the qualitative assembly pathway, which maintains the characteristic progression from linear chains to ring-containing oligomers to interconnected networks. \par
The computational efficiency of nb-CG-ZIF-FF is particularly significant because it opens doors to investigating systems that remain prohibitively expensive for all-atom simulations. This capability allows us to probe low-concentration regimes that closely mimic experimental synthesis conditions and to systematically modulate the metal-to-ligand ratio, a key synthetic parameter that critically governs polymorph selection, defect generation, and crystal morphology.\cite{luczak2023morphology,Talosig2024}
The ability to computationally screen different synthesis compositions at realistic scales provides a powerful complement to experimental efforts dedicated to MOF rational design. Looking ahead, we are interested in refining the model to capture the transition from amorphous to crystalline phases and in using it to help design better synthesis protocols for making MOFs with specific desired properties.

\section*{Conflicts of interest}
There are no conflicts of interest to declare.

\section*{Data availability}

The data supporting this article have been included as part of the Supplementary Information or in github at \url{https://github.com/rosemino/MAGNIFY/tree/main/nb-CG-ZIF-FF_development} .

\begin{acknowledgments}
This work was funded by the European Union ERC Starting grant MAGNIFY, grant number 101042514. This work was granted access to the HPC resources of the MeSU platform at Sorbonne Université, where the simulations were performed.
\end{acknowledgments}

\section{Supporting Information}
\noindent\textbf{A. DMSO Incorporation into Crystalline ZIF-8} \\

To investigate the equilibrium uptake of dimethylsulfoxide (DMSO) solvent molecules within the ZIF-8 framework, we perform hybrid Monte Carlo (MC)/ molecular dynamics (MD) simulations via LAMMPS.\cite{Thompson2022} Within this hybrid approach, MC and MD simulations are alternated: every 500 MD simulation steps, 100 insertions/deletions of DMSO molecules are attempted within the ZIF-8 framework. During each MC attempt, the algorithm randomly tries to either insert a new DMSO molecule or delete an existing one, with acceptance determined by the Metropolis criterion. The MC algorithm used concerns that of the Grand-Canonical ensemble.\cite{frenkel2023understanding} This hybrid approach allows the system to relax structurally through MD while equilibrating DMSO loading through MC moves. Note that we keep both the MOF and DMSO molecules flexible during the process.

\begin{figure}[h]
\centering
\includegraphics[width=0.6\linewidth]{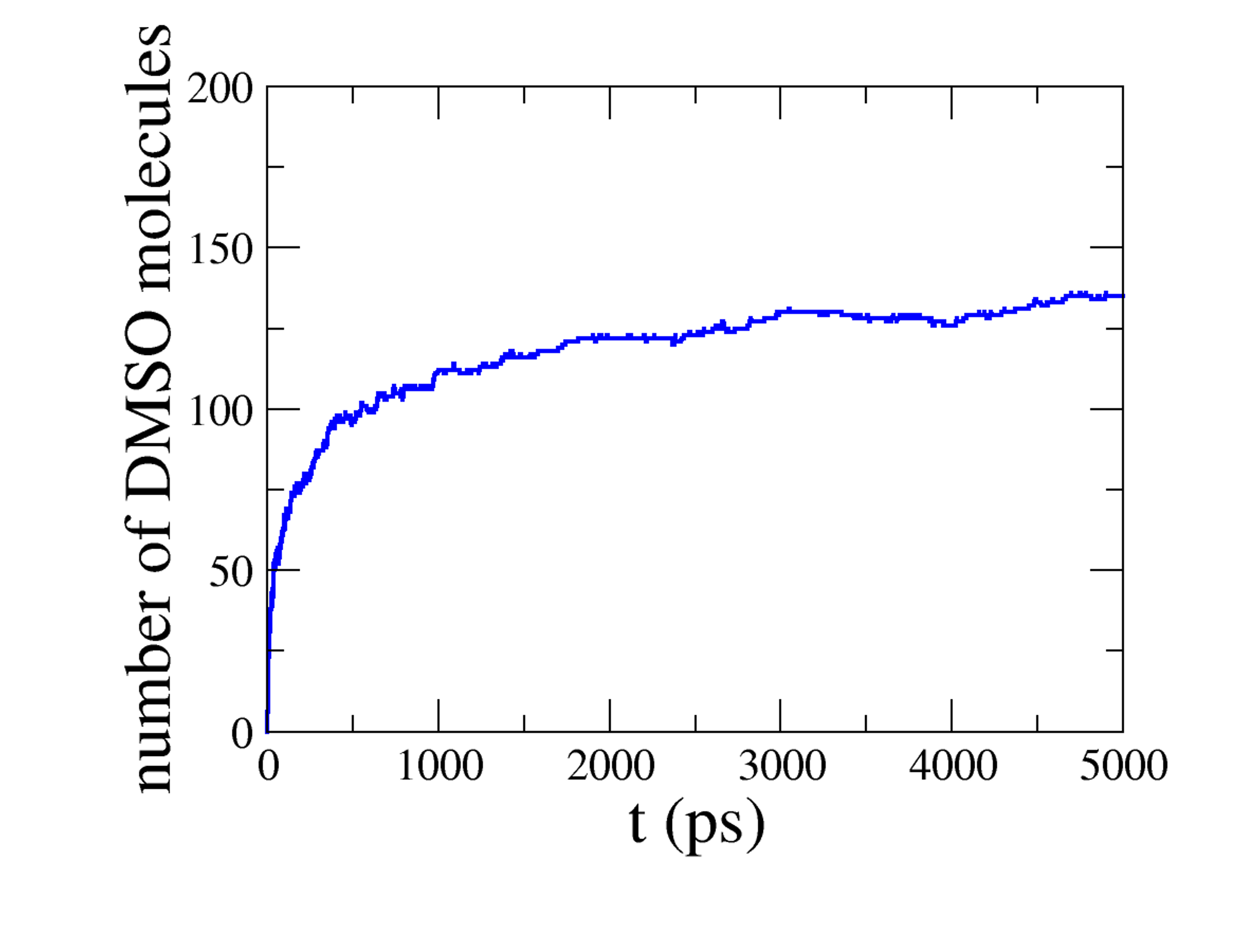}
\caption{Time evolution of DMSO uptake in the ZIF-8 framework.}
\label{fig:s1}
\end{figure}

In Fig. \ref{fig:s1}, we plot the number of DMSO molecules inside the ZIF-8 framework as a function of simulation time, where the timestep corresponds to 0.001 fs. We choose a low timestep to avoid numerical instabilities that could occur due to molecule insertion. The system exhibits an initial rapid uptake phase as DMSO molecules are inserted into the accessible pore space, followed by equilibration to a plateau value of approximately 135 molecules for the whole simulation box, indicating saturation. The saturated loading represents the thermodynamic equilibrium between the framework-adsorbed DMSO and a virtual reservoir at a chemical potential corresponding to liquid DMSO.
Further, to confirm that the DMSO loading represents complete pore saturation, we perform a series of GCMC simulations at different reservoir pressures while maintaining T=298 K. Table \ref{tab:tables1} gives the number of equilibrium DMSO loadings per pore obtained at reservoir pressures ranging from 10 to 80 atm. No significant variation is observed across the pressure range studied. We thus decided to keep the lower P value tested.

\begin{table}[t]
\centering
\caption{DMSO loading in ZIF-8 across a range of GCMC virtual reservoir pressures}
\label{tab:tables1}
\begin{tabular}{lc}
\hline
Pressure (atm) & number of DMSO molecules per pore\\
\hline
10       & 14.38 \\
30      & 15.75 \\
50     & 15.85 \\
80     & 16.10 \\
\hline
\end{tabular}
\end{table}

\vspace{0.5 cm}

\noindent\textbf{B. Fitting Details of MS-CG–Derived Potentials}\\

\begin{figure*}[h]
\centering 
\includegraphics[width=1.0\linewidth]{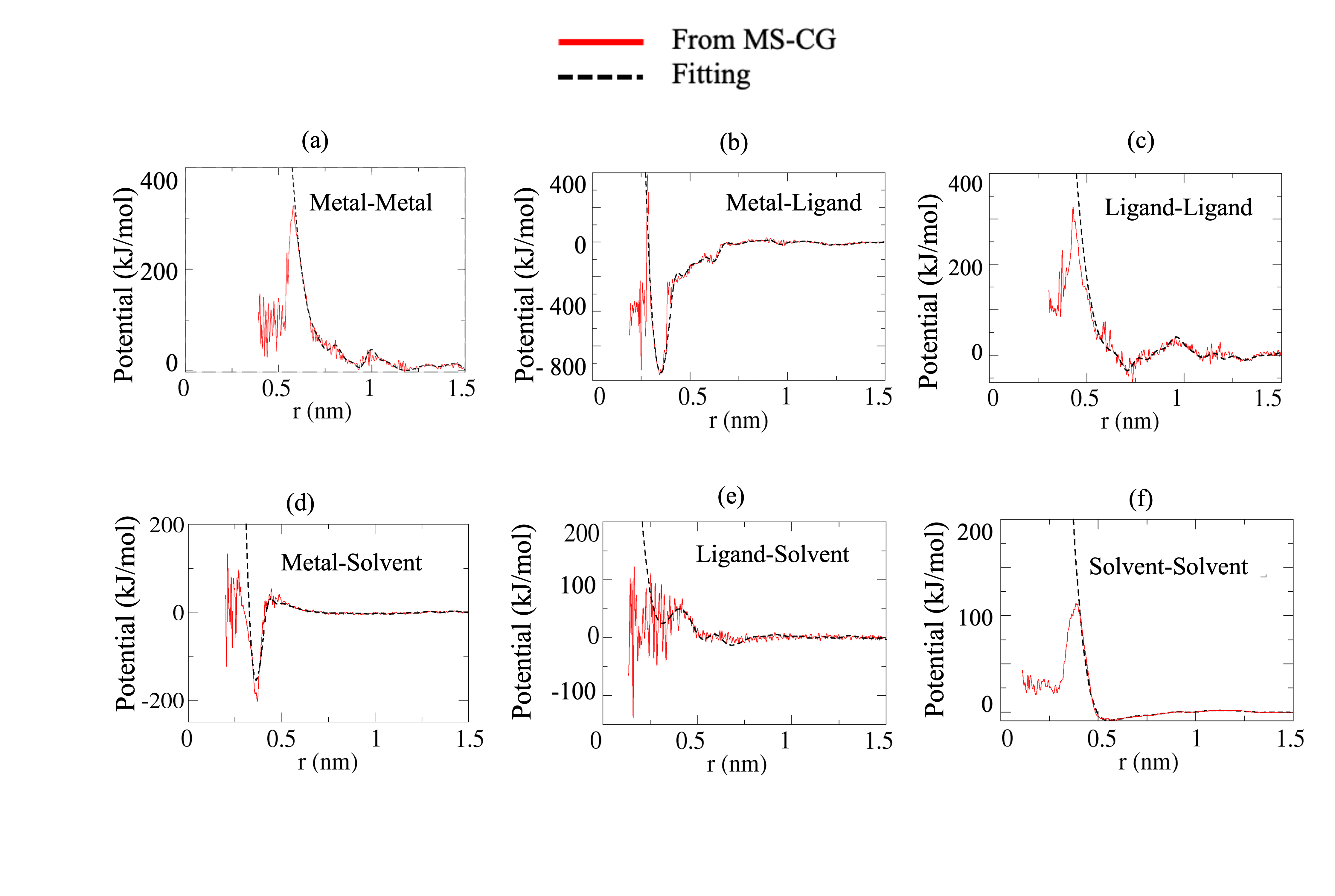}
\caption{Pair potentials obtained from MS-CG (red) along with the post-processing fitting (dashed black line).}
\label{fig:s2}
\end{figure*}

Fig. \ref{fig:s2}. shows the pair potentials obtained after the MS-CG procedure for the ZIF-8 system. Panels (a–f) represent the metal–metal, metal–ligand, ligand–ligand, metal–solvent, ligand–solvent, and solvent–solvent interaction potentials, respectively. Red lines show the raw output of forces from the force matching procedure, while black dashed lines indicate the fitted curves after smoothing and extrapolation. The raw data exhibit significant statistical noise due to finite sampling and thermal fluctuations inherent to the molecular dynamics trajectories used in the MS-CG procedure. To obtain smooth, continuous potentials suitable for CG simulations, these data are processed using cubic spline fits, which effectively reduce noise while preserving the essential features of the interaction profiles. The upper boundary of the sampling range is uniformly set to 1.5 nm for all pair potentials, which is sufficient to capture long-range behavior. However, determination of the lower boundary limits requires careful consideration and is a critical aspect of the force field development process. \cite{Alvares2024} As discussed in the main manuscript, developing a force field that simultaneously stabilizes the crystalline ZIF-8 structure while capturing realistic self-assembly dynamics presents a significant challenge and necessitates extensive manual refinement of both the trajectory weighting scheme and the lower boundary distances used in force matching. After smoothening, fitting and extrapolating both the sampled and non sampled region, we validate the obtained potentials by checking their ability to reproduce both structural and dynamic properties. In cases where potentials are excessively repulsive, the crystalline structural properties—including lattice parameters and radial distribution functions—were accurately reproduced. However, these excessive repulsive regions of the potentials at low pair distances prevented the system from exploring the transient, high-energy configurations necessary for bond formation and rearrangement, severely hindering self-assembly processes such as metal-ligand coordination, cluster formation and nucleation.
Conversely, in cases where the potentials are less restrictive, the system exhibits more dynamic behavior and successfully captures early-stage self-assembly, including cluster formation and metal-ligand coordination events. However, these softer potentials compromised the structural integrity of the crystalline phase, resulting in distorted lattice parameters, incorrect coordination geometries, or even structural collapse during extended equilibration simulations. In addition to the lower boundary limits, the relative contribution of crystalline and self-assembly trajectories also influence the resulting potentials. To resolve this, we employ an iterative optimization strategy that systematically tested force fields against a comprehensive set of validation criteria spanning both structural and dynamical properties. This included: lattice parameters (a, b, c), radial distribution functions g(r) for metal-metal, metal–ligand and ligand–ligand pairs, coordination numbers, time evolution of coordination number and ring distribution. Through systematic variation of both the trajectory weighting and the lower boundary cutoff distances (which determine the minimum sampled distances and thus influence the steepness of the repulsive wall), we identified an optimal balance. The final optimized lower boundary values, listed in Table \ref{tab:tables2}, represent the minimum distances used for the fitting and define the onset of the repulsive region in each pair potential. The resulting force field successfully reproduces both the thermodynamic stability of the crystalline ZIF-8 framework and the kinetic pathways of solvothermal self-assembly.
All force field files and associated scripts are available at the following GitHub repository: \url{https://github.com/rosemino/MAGNIFY/tree/main/nb-CG-ZIF-FF_development}.

\begin{table}[t]
\centering
\caption{Final optimized lower boundary values for the pair potential}
\label{tab:tables2}
\begin{tabular}{lc}
\hline
Interaction type & Lower repulsive boundary (nm) \\
\hline
Metal--Metal       & 0.621 \\
Metal--Ligand      & 0.295 \\
Ligand--Ligand     & 0.560 \\
Metal--Solvent     & 0.344 \\
Ligand--Solvent    & 0.264 \\
Solvent--Solvent   & 0.412 \\
\hline
\end{tabular}
\end{table}

\vspace{0.5 cm}

\noindent\textbf{C.Details of coordination number calculation and ring analyses}\\

\textbf{Coordination Number Calculation:} To quantify the structural evolution during ZIF-8 formation, we calculate the coordination number (CN) for each metal center at every timestep throughout the simulation trajectory. CN represents the number of ligands directly bound to a given metal center and serves as a key indicator of framework connectivity and self-assembly progress.
The coordination cutoff distance was set to 3.5 \AA, corresponding to the first minimum (valley) following the first peak in the metal-ligand radial distribution function obtained from the crystalline ZIF-8 trajectory. This cutoff ensures that only ligands within the first coordination shell are counted, excluding more distant neighbors and preventing spurious coordination assignments. The coordination number calculation is implemented as a post-processing analysis tool written in Fortran 90. The algorithm processes LAMMPS trajectory files containing metal(96 Zn ions) and ligand (192 2-methylimidazolate ions) coordinates at each simulation step. We perform this analysis over 5000 trajectory frames with data saved every 0.1 ps, corresponding to 500 ps of total simulation time. For each frame, metal centers are classified according to their coordination number (CN = 0 to 4), and the population of each coordination state is recorded. CN = 0 corresponds to isolated, uncoordinated metal centers, while CN = 4 represents fully coordinated metal centers with tetrahedral geometry as in crystalline ZIF-8. Intermediate coordination states (CN = 1, 2, 3) represent various oligomeric species and partially assembled framework fragments. \par
\textbf{Ring Analyses:} To characterize the topology and connectivity of assembled structures during ZIF-8 formation, we perform ring analyses to identify closed metal-ligand rings of various sizes. Ring structures are fundamental topological features of ZIF-8 frameworks, and their emergence during assembly provides insight into the nucleation and growth mechanisms. A ring is defined as a closed cyclic path consisting of alternating metal and ligand beads, where each ligand in the ring bridges exactly two metal centers that are adjacent in the cycle. We focused on rings containing 3 to 8 metal centers, as these sizes are most relevant to ZIF-8 structure (which contains 4- and 6-membered rings as primary building blocks of the sodalite topology).
The ring analyses were implemented in Python using the NetworkX graph theory library. \cite{osti_960616} The algorithm constructs a graph representation where metal centers serve as nodes and edges are added between metal pairs that share a bridging ligand. The procedure operates as follows. For each ligand, we determine which metal centers lie within a coordination cutoff distance of 3.5 \AA. Two metals are considered as neighbors if the distance between them is less than 6.6 \AA  and if a ligand bridges them (ensuring they are bonded through the ligand rather than accidentally close in space). An edge is then added to the graph between these two metal centers. Once the metal connectivity graph is constructed, we use NetworkX's minimum cycle basis algorithm to identify all independent rings in the graph. This algorithm finds the smallest set of fundamental cycles from which all other cycles can be constructed through linear combinations. The resulting rings are filtered to retain only those with 3 to 8 members. At each timestep, the algorithm outputs the counts of validated rings for each size. These counts are written to a time series file showing the temporal evolution of ring populations throughout the self-assembly process. The emergence and evolution of ring populations is discussed in the main manuscript. CN calculation and Ring Analysis scripts are available in the following GitHub repository: \url{https://github.com/rosemino/MAGNIFY/tree/main/nb-CG-ZIF-FF_development} and \url{https://github.com/rosemino/MAGNIFY/tree/main/ZIF-8_nucleation_T_concentration_solvent}, respectively.

\bibliography{arXiv}


\end{document}